\documentstyle[aaspp4]{article}

\lefthead{McClure}
\righthead{The Subgiant CH Stars}

\begin{document}

\title{The Binary Nature of the Subgiant CH Stars}

\author{Robert D. McClure}
\affil{Dominion Astrophysical Observatory,
    Herzberg Institute of Astrophysics,\\
    National Research Council, Canada,
    5071 W. Saanich Road, Victoria, BC, V0S 1N0\\
    email:  mcclure@dao.nrc.ca}

\begin{abstract}

Repeated radial velocities have been measured for a sample of 10
subgiant CH (sgCH) stars over a period of $\sim$15 years.  Long-term
velocity variations are exhibited by all but one star, and
spectroscopic orbits have been calculated for six of them.  The periods
are long, ranging from 876 to 4144 days.  The distribution of
eccentricities and mass functions for sgCH star orbits are similar to
those for giant barium and CH stars, exhibiting significant orbital
dissipation relative to normal late-type binaries.  It is concluded
that all sgCH stars are binaries, and like the barium and CH stars,
they have had mass transferred from a former asymptotic giant-branch
star.

\end{abstract}


\section{Introduction}

From examination of objective prism spectra, Bond (1974) discovered a
class of G-type stars with enhanced features of CH and s process
elements in their spectra, and with absolute magnitudes placing them
near and above the main sequence.  With relatively high velocities and
weak metal lines, they appeared to be Population II objects and they
were referred as subgiant CH (sgCH) stars, because he suggested that
they would eventually become classical CH stars when they evolved away
from the main sequence and up the giant branch.  McClure (1984a, 1985)
has reviewed status of the barium, CH and related stars, including the
sgCH stars.  Abundance analyses have been done by Sneden and Bond
(1976), Luck and Bond (1982), Sneden (1983), Krishnaswamy and Sneden
(1985), Smith and Lambert (1986), Luck and Bond (1991), and Smith
Coleman and Lambert (1993).  These analyses of a larger sample of stars
indicated that most sgCH stars are moderately metal deficient with
[Fe/H] = $-0.1$ to $-0.8$, and with s process enhancements similar to
the barium stars.  The classical CH stars have higher overabundances of
the heavy s process elements (Ba through Sm), and C$_2$ bands
(indicating C/O $>$ 1) which are lacking in the sgCH stars.  The
velocities of most of the sgCH stars are also moderate, being more like
the old disk than the halo.  The essential conclusion of these papers
is that despite the name which was originally assigned to them, the
sgCH stars are probably the progenitors of moderately metal-deficient
barium stars rather than the halo CH stars themselves.  A low Li
abundance in the sgCH stars was thought to be a problem with their role
as progenitors of the barium or CH stars as pointed out by Smith and
Lambert (1986), but Lambert, Smith and Heath (1993) have dismissed this
as being due to contamination of the Li line by an unidentified line.

The terminology for the sgCH stars has become confusing as a result of
the findings of only moderate metal-deficiency.  They tend to be referred
to in the literature as ``dwarf barium stars''  ( e.g. Jorissen and
Boffin 1992, North, Jorissen and Mayor 1996), and overlap with stars
referred to as ``F Str $\lambda4077$'' stars (North and Duquennoy
1991).  Han et al. (1995) refer to ``pre-Ba/CH'' stars which undoubtedly
includes the sgCH stars.

Regardless of their role as progenitors of the barium stars or of the
CH stars, the frequency of binaries among the sgCH stars is of very
significant interest.  Both the barium and CH stars have been shown to
be long period binaries, and their peculiar abundances are likely the
result of mass transfer from a former asymptotic giant-branch (AGB)
star whose atmosphere has been contaminated from products of helium
shell-flashing (McClure, Fletcher and Nemec 1980; McClure 1983; McClure
1984b; Jorissen and Mayor 1988; Webbink 1986; McClure and Woodsworth
1990).   Smith and Demarque (1980) first suggested that the sgCH stars
may be mass-transfer binaries since they found it impossible to explain
them by internal mixing of helium core-flash material.  A program was
begun by the present author, therefore,  in the early 1980's, to
monitor radial velocities of sgCH stars.  Some very preliminary results
from this program have been mentioned in the literature (McClure 1985,
1989) indicating a high binary frequency, and Luck and Bond (1991) also
observed several sgCH stars which appeared to exhibit variable
velocity.  As a result, most authors have assumed over the last few
years that all sgCH stars are binaries.  The evidence from these
preliminary data, however, is somewhat meager.  The present paper
presents more complete velocity data for this sample of stars, and sets
the binary frequency for the sgCH stars on a firm basis.  Results of
similar velocity monitoring of sgCH stars has recently been mentioned
in a conference paper presented by North, Jorissen, and Mayor (1996).

\section{Radial Velocity Observations}

The radial-velocity observations were made at the Dominion
Astrophysical Observatory in Victoria, using the radial-velocity
spectrometer at the coud\'e focus of the 1.2 meter telescope.  See
Fletcher et al.  (1982) and McClure et al. (1985) for a description of
the instrument.  It is capable of a precision of about $\pm0.3$ km
s$^{-1}$, although with the limited integration time for each
observation, the observational error for the sgCH stars is closer to
double that value.  The stellar sample was taken from Bond (1974),
supplemented by a few stars from Bond (private communication) that were
analyzed later by Luck and Bond (1991).  One of these, HD122202, is
representative of an earlier class of mid F-type stars with strong Sr,
discussed by Bidelman (1981), and more recently by North and Duquennoy
(1991).  They were referred to by these authors as F str $\lambda4077$
stars, but they are almost certainly just earlier examples of G-type
sgCH stars.

For most of the stars in this sample Luck and Bond (1991) list radial
velocities measured from the same coud\'e plates from which they analyzed
abundances.  These velocities are of comparable precision and can be
combined with the present data after a small systematic velocity shift
of $-0.8$ km s$^{-1}$ (see below).  The radial-velocities and
Julian dates ($-2400000$ days) for each sgCH star are listed in Table 1.
Those few entries with Julian dates earlier than $2444000$
and marked with an asterisk are from Luck and Bond (1991).


The velocity data are plotted versus Julian date in the upper part of
Figure 1.  The time span exhibited in these plots is nearly 25 years.
Considering that the errors of velocities are approximately the size of
the plotted symbols, it is apparent that at least eight of the ten
stars exhibit velocity variations representative of long-term binary
motion.  One further star HD 4395 exhibits some evidence of variable
velocity, although the amplitude is small, and no convincing binary
orbit could be calculated.  The binary frequency is consistent,
therefore, with all sgCH stars being binaries, with orbits viewed at
random orientations.  Luck and Bond (1991) also suggested that three of
their stars are velocity variables, although one of these, HD 88446, is
the one star in the present analysis that does not appear to exhibit
long term variations.


\section{Binary Star Orbits}

Orbits have been computed for six of the ten sgCH stars, although for
those for which the period is comparable to the time span of the
observations the orbits should be considered somewhat preliminary.
Because the velocity variations for the sgCH stars are very small, one
must be careful about systematic differences when combining data from a
different program and different instruments.  See the discussion, for
example, on standard-velocity zero-pints by IAU Commission 30 (Andersen
1990), where uncertainties approaching 1 km s$^{-1}$ are discussed.
The orbits were first calculated, therefore, using only the data from
the present set of observations, giving the velocities from Luck and
Bond (1991) zero weight.  Residuals from the velocity curves for their
observations were calculated and a systematic difference of $+0.85$ km
s$^{-1}$ was found with a mean error of $\pm0.38$ km s$^{-1}$.  The
apparently constant velocity star HD 88446 also gives a similar
positive error ($+1.18$  km s$^{-1}$) with respect to the present data
for that star, although this residual was not included in the average
because a long term variation for the star cannot be ruled out.  The
Luck and Bond data were all shifted by $-0.8$ km s$^{-1}$, therefore,
and the orbits were calculated again, giving all observations unit
weight.  The resulting average residual for the Luck and Bond data from
the final orbital calculations is $+0.04\pm0.15$ km s$^{-1}$.  The
orbital elements for the six stars are listed in Table 2.  The
velocities as a function of phase, resulting from these solutions are
plotted in the lower part of Figure 1, and phases and residuals from
the orbital curves are listed for the individual observations in Table
1.

Orbit calculations were attempted for one further star HD 182274, but a
unique orbit could not be determined.  An orbit with low eccentricity
and a period of 8000 days or more is possible, but the amplitude and
period are uncertain.  It is also possible that the orbit might have
very high eccentricity, with the maximum velocity occurring immediately
before the present observations were begun, but it could be several
more years before this could be confirmed or ruled out.


The cumulative distribution of eccentricities and mass functions are
shown in Figure 2 for the sgCH stars along with similar distributions
for the barium and CH stars and for normal K giants; the latter
distributions are taken from McClure and Woodsworth (1990).  In both
the eccentricity and mass functions, the sgCH star distributions lie
between those for the barium and CH stars.  Relative to normal K
giants there is evidence, from the lower eccentricities, for orbital
dissipation in the sgCH sample, and also for a much more homogeneous
distribution of mass functions.  McClure and Woodsworth (1990) showed
that the barium and CH star mass-function distributions could be
modeled with stars of approximately uniform secondary mass,
whereas that for K giants required a large range in mass.  They found,
assuming reasonable masses for the barium and CH primaries,
that in both cases the secondary masses are near
0.6 solar masses, typical of white dwarfs.  The same must be
true for the sgCH stars, given that their mass functions
are so similar to those of the barium stars.
The distributions exhibited
in Figure 2 lend support to the suggestion of Luck and Bond (1991)
based on chemical abundances, that the sgCH stars may be precursors to
a moderately metal-poor population of giant barium stars.  The orbital
dissipation supports the hypothesis that these stars have undergone
mass exchange.


\subsection{Summary} \label{nstars}

Radial velocity measurements over the last $\sim15$ years, have shown
that the sgCH stars are likely all binaries.  A similar conclusion has
recently been suggested in a conference paper by North, Jorissen and
Mayor (1996), who have found a high binary frequency among what they
refer to as barium dwarfs, which include sgCH stars as well as earlier
F-type stars with strong lines of s process elements.  The orbital
eccentricities and mass functions for sgCH stars are distributed in a
very similar way to those of the barium and CH stars.  It is likely
that the sgCH stars are precursors to the barium stars as suggested by
Luck and Bond (1991), their peculiar abundances being derived from mass
transfer of material contaminated through helium shell-flashing in an
AGB star that has since evolved to become an unseen white-dwarf.

\acknowledgments

I wish to thank Murray Fletcher, Les Saddlemyer, Doug Bond, and Frank
Younger for help over the years in operation of the radial-velocity
spectrometer.  I thank Howard Bond for providing me with a list of sgCH
stars many years ago prior to publication.

\def\nd{\nodata}
\def\ts{\extracolsep{-.2em}}
\def\s{\extracolsep{-.4em}}
\def\ss{\extracolsep{-.7em}}
\def\sss{\extracolsep{-1.em}}

\textheight=8.3in
\footnotesize

\begin{deluxetable}{@{\hspace{-.2em}}c@{\ts}c@{\s}c@{\s}c@{\s}c@{\sss}c@{\s}c@{\s}c@{\s}
c@{\s}c@{\sss}c@{\s}c@{\s}c@{\s}c}
\tablenum{1}
\tablewidth{500pt}
\tablecaption{Radial-Velocity Data \& Ephemerides for Orbital Solutions}
\tablehead{
\colhead{JD}             & \colhead{\phs RV}             &
\colhead{\phn\phn O-C}         & \colhead{Phase}             &
\colhead{}              &
\colhead{JD}             & \colhead{\phs RV}             &
\colhead{\phn\phn O-C}         & \colhead{Phase}             &
\colhead{}              &
\colhead{JD}             & \colhead{\phs RV}             &
\colhead{\phn\phn O-C}         & \colhead{Phase\phn} \\
\colhead{\hspace{-.2em}$-2400000$}                 & 
\multicolumn{2}{c}{\phd\phn km s$^{-1}$}    & \colhead{} &
\colhead{}                 & \colhead{$-2400000$}        &
\multicolumn{2}{c}{\phd\phn km s$^{-1}$}          &
\colhead{}                 &
\colhead{}                 & \colhead{$-2400000$}        &
\multicolumn{2}{c}{\phd\phn km s$^{-1}$}          &
\colhead{}}

\startdata

 \multicolumn{4}{c}{\bf HD 4395     }               &&  \multicolumn{4}{c}{\bf HD 88446    }               &&  \multicolumn{4}{c}{\bf HD 122202   }               \nl
43382.978 &\phd\phd$     -1.6^{\displaystyle*} $&\phs  ....   &  ....   && 45849.770 &\phs      60.99  &\phs  ....   &  ....   && 45713.093 &$        -12.20 $&\phs   0.75  &   0.007 \nl
43385.951 &\phn$     -0.90 $&\phs  ....   &  ....   && 46852.046 &\phs      61.10  &\phs  ....   &  ....   && 45752.861 &$        -12.05 $&\phs   0.19  &   0.038 \nl
45384.648 &\phn$     -2.39 $&\phs  ....   &  ....   && 47087.048 &\phs      60.45  &\phs  ....   &  ....   && 45803.886 &$        -10.64 $&\phs   0.63  &   0.077 \nl
45583.938 &\phn$     -2.10 $&\phs  ....   &  ....   && 47141.989 &\phs      61.06  &\phs  ....   &  ....   && 45849.793 &$        -10.86 $&$     -0.43 $&   0.113 \nl
45667.750 &\phn$     -2.63 $&\phs  ....   &  ....   && 47223.879 &\phs      60.72  &\phs  ....   &  ....   && 46234.875 &\phn$     -7.64 $&\phs   0.79  &   0.411 \nl
45712.590 &\phn$     -3.16 $&\phs  ....   &  ....   && 48019.722 &\phs      61.28  &\phs  ....   &  ....   && 46875.983 &$        -14.81 $&$     -0.34 $&   0.908 \nl
45938.930 &\phn$     -4.51 $&\phs  ....   &  ....   && 48387.754 &\phs      61.61  &\phs  ....   &  ....   && 47223.943 &\phn$     -8.58 $&\phs   0.53  &   0.178 \nl
46377.832 &\phn$     -1.31 $&\phs  ....   &  ....   && 50123.860 &\phs      60.81  &\phs  ....   &  ....   && 47260.961 &\phn$     -9.44 $&$     -0.76 $&   0.207 \nl
46657.941 &\phn$     -0.54 $&\phs  ....   &  ....   && 50127.798 &\phs      61.30  &\phs  ....   &  ....   && 47316.787 &\phn$     -7.83 $&\phs   0.39  &   0.250 \nl
46717.785 &\phn$     -0.77 $&\phs  ....   &  ....   && 50128.893 &\phs      61.04  &\phs  ....   &  ....   && 48019.750 &$        -14.56 $&$     -0.25 $&   0.795 \nl
47031.891 &\phs\phn   1.41  &\phs  ....   &  ....   && 50129.830 &\phs      59.57  &\phs  ....   &  ....   && 48387.790 &$        -11.44 $&$     -0.24 $&   0.080 \nl
47116.793 &\phn$     -0.74 $&\phs  ....   &  ....   && 50142.816 &\phs      60.35  &\phs  ....   &  ....   && 49560.771 &$        -13.87 $&$     -0.56 $&   0.990 \nl
47763.855 &\phn$     -1.03 $&\phs  ....   &  ....   &&           &                 &             &         && 50130.019 &\phn$     -9.63 $&$     -1.00 $&   0.431 \nl
47872.645 &\phn$     -0.96 $&\phs  ....   &  ....   &&  \multicolumn{4}{c}{\bf HD 89948    }               && 50142.991 &\phn$     -8.55 $&\phs   0.19  &   0.441 \nl
48232.711 &\phs\phn   0.25  &\phs  ....   &  ....   && 43201.875 &\phd\phd\phd      18.0$^{\displaystyle*}$  &\phs  ....   &  ....   &&           &                 &             &         \nl
49261.852 &\phn$     -1.27 $&\phs  ....   &  ....   && 43202.795 &\phd\phd\phd      18.8$^{\displaystyle*}$  &\phs  ....   &  ....   &&  \multicolumn{4}{c}{\bf HD 182274   }               \nl
49647.723 &\phn$     -0.73 $&\phs  ....   &  ....   && 45313.039 &\phs      22.00  &\phs  ....   &  ....   && 42675.698 &$        -13.9^{\displaystyle*} $&\phs  ....   &  ....   \nl
50364.834 &\phn$     -0.81 $&\phs  ....   &  ....   && 46903.774 &\phs\phn   5.58  &\phs  ....   &  ....   && 42676.657 &$        -14.3^{\displaystyle*} $&\phs  ....   &  ....   \nl
50388.804 &\phn$     -0.42 $&\phs  ....   &  ....   && 47142.018 &\phs      15.08  &\phs  ....   &  ....   && 42999.753 &$        -12.6^{\displaystyle*} $&\phs  ....   &  ....   \nl
          &                 &             &         && 47223.893 &\phs      16.10  &\phs  ....   &  ....   && 45178.719 &$        -13.88 $&\phs  ....   &  ....   \nl
 \multicolumn{4}{c}{\bf HD 11377    }               && 50127.889 &\phs      17.99  &\phs  ....   &  ....   && 45205.742 &$        -15.18 $&\phs  ....   &  ....   \nl
43384.966 &$        -26.4^{\displaystyle*} $&$     -0.02 $&   0.552 && 50129.861 &\phs      18.37  &\phs  ....   &  ....   && 45284.637 &$        -16.97 $&\phs  ....   &  ....   \nl
45667.773 &$        -27.37 $&$     -0.09 $&   0.103 && 50142.846 &\phs      17.54  &\phs  ....   &  ....   && 45452.949 &$        -18.77 $&\phs  ....   &  ....   \nl
46335.001 &$        -28.94 $&\phs   0.55  &   0.265 &&           &                 &             &         && 45583.711 &$        -18.49 $&\phs  ....   &  ....   \nl
46377.854 &$        -30.50 $&$     -1.01 $&   0.275 &&  \multicolumn{4}{c}{\bf BD+17 2537  }               && 45667.574 &$        -17.49 $&\phs  ....   &  ....   \nl
46704.944 &$        -29.02 $&\phs   0.09  &   0.354 && 43242.906 &\phn\phn\phd   4.6$^{\displaystyle*}$  &$     -0.10 $&   0.303 && 45803.991 &$        -17.68 $&\phs  ....   &  ....   \nl
46802.745 &$        -28.12 $&\phs   0.77  &   0.378 && 45367.945 &\phs\phn   6.67  &$     -0.65 $&   0.486 && 45915.778 &$        -19.43 $&\phs  ....   &  ....   \nl
47031.901 &$        -28.52 $&$     -0.28 $&   0.433 && 45368.059 &\phs\phn   6.69  &$     -0.63 $&   0.486 && 46040.628 &$        -18.22 $&\phs  ....   &  ....   \nl
47086.798 &$        -29.23 $&$     -1.16 $&   0.446 && 45452.797 &\phs\phn   7.18  &\phs   0.12  &   0.533 && 46334.758 &$        -18.09 $&\phs  ....   &  ....   \nl
47789.967 &$        -24.63 $&\phs   0.58  &   0.616 && 45713.030 &\phs\phn   4.43  &\phs   0.42  &   0.678 && 46657.844 &$        -19.46 $&\phs  ....   &  ....   \nl
47872.743 &$        -24.14 $&\phs   0.70  &   0.636 && 45752.855 &\phs\phn   3.56  &\phs   0.31  &   0.700 && 46698.664 &$        -19.03 $&\phs  ....   &  ....   \nl
48232.772 &$        -24.02 $&$     -0.78 $&   0.723 && 45803.873 &\phs\phn   1.51  &$     -0.66 $&   0.729 && 46717.659 &$        -18.76 $&\phs  ....   &  ....   \nl
49261.977 &$        -22.88 $&\phs   0.26  &   0.972 && 45849.775 &\phs\phn   1.03  &$     -0.08 $&   0.754 && 46970.873 &$        -19.55 $&\phs  ....   &  ....   \nl
49647.870 &$        -26.32 $&$     -0.22 $&   0.065 && 46234.865 &\phn$     -7.45 $&$     -0.52 $&   0.969 && 46990.781 &$        -19.53 $&\phs  ....   &  ....   \nl
50129.639 &$        -29.20 $&$     -0.29 $&   0.181 && 46875.962 &\phs\phn   5.96  &\phs   0.61  &   0.326 && 47005.941 &$        -20.77 $&\phs  ....   &  ....   \nl
50364.845 &$        -28.96 $&\phs   0.46  &   0.238 && 47141.945 &\phs\phn   7.47  &\phs   0.15  &   0.474 && 47086.803 &$        -19.41 $&\phs  ....   &  ....   \nl
50388.812 &$        -29.01 $&\phs   0.43  &   0.244 && 47223.930 &\phs\phn   7.21  &\phs   0.04  &   0.519 && 47317.918 &$        -19.47 $&\phs  ....   &  ....   \nl
          &                 &             &         && 47260.945 &\phs\phn   7.12  &\phs   0.13  &   0.540 && 47377.791 &$        -18.94 $&\phs  ....   &  ....   \nl
 \multicolumn{4}{c}{\bf HD 88446    }               && 47316.768 &\phs\phn   7.02  &\phs   0.43  &   0.571 && 47789.737 &$        -19.20 $&\phs  ....   &  ....   \nl
43147.950 &\phd\phd\phd      60.9$^{\displaystyle*}$  &\phs  ....   &  ....   && 48019.740 &\phn$     -6.25 $&\phs   0.62  &   0.962 && 48019.984 &$        -18.74 $&\phs  ....   &  ....   \nl
43200.729 &\phs      59.70  &\phs  ....   &  ....   && 48387.766 &\phn$     -1.31 $&$     -0.40 $&   0.167 && 48089.833 &$        -19.64 $&\phs  ....   &  ....   \nl
43203.851 &\phs      63.40  &\phs  ....   &  ....   && 49560.747 &\phn$     -1.92 $&$     -0.02 $&   0.821 && 48162.694 &$        -19.06 $&\phs  ....   &  ....   \nl
45053.863 &\phs      61.92  &\phs  ....   &  ....   && 50128.943 &\phn$     -2.72 $&$     -0.31 $&   0.137 && 49175.894 &$        -17.40 $&\phs  ....   &  ....   \nl
45116.715 &\phs      60.60  &\phs  ....   &  ....   && 50142.930 &\phn$     -1.51 $&\phs   0.52  &   0.145 && 49261.682 &$        -16.82 $&\phs  ....   &  ....   \nl
45284.039 &\phs      62.33  &\phs  ....   &  ....   &&           &                 &             &         && 49560.788 &$        -16.11 $&\phs  ....   &  ....   \nl
45313.020 &\phs      61.20  &\phs  ....   &  ....   &&  \multicolumn{4}{c}{\bf HD 122202   }               && 49647.656 &$        -17.39 $&\phs  ....   &  ....   \nl
45367.936 &\phs      60.14  &\phs  ....   &  ....   && 43240.847 &$        -11.3^{\displaystyle*} $&$     -0.34 $&   0.090 && 50124.083 &$        -15.76 $&\phs  ....   &  ....   \nl
45410.867 &\phs      61.05  &\phs  ....   &  ....   && 45368.043 &$        -13.94 $&$     -0.31 $&   0.739 && 50129.095 &$        -15.04 $&\phs  ....   &  ....   \nl
45712.887 &\phs      60.80  &\phs  ....   &  ....   && 45411.039 &$        -13.52 $&\phs   0.55  &   0.773 && 50364.696 &$        -14.51 $&\phs  ....   &  ....   \nl
45803.867 &\phs      60.91  &\phs  ....   &  ....   && 45452.871 &$        -14.19 $&\phs   0.21  &   0.805 && 50388.590 &$        -14.12 $&\phs  ....   &  ....   \nl

 \multicolumn{4}{c}{\bf HD 202020   }               &&  \multicolumn{4}{c}{\bf HD 204613   }               &&  \multicolumn{4}{c}{\bf HD 216219   }               \nl
43055.722 &$        -21.5^{\displaystyle*} $&\phs   0.48  &   0.030 && 46704.710 &$        -93.09 $&$     -0.41 $&   0.118 && 44428.938 &\phn$     -4.13 $&\phs   0.06  &   0.645 \nl
45583.844 &$        -18.52 $&$     -0.15 $&   0.255 && 46717.710 &$        -92.40 $&$     -0.06 $&   0.133 && 44485.801 &\phn$     -4.48 $&$     -0.15 $&   0.660 \nl
45612.766 &$        -18.26 $&\phs   0.32  &   0.269 && 46970.918 &$        -88.15 $&$     -0.02 $&   0.421 && 44606.641 &\phn$     -5.29 $&$     -0.60 $&   0.691 \nl
45667.582 &$        -18.73 $&\phs   0.36  &   0.295 && 46980.960 &$        -89.06 $&$     -0.96 $&   0.433 && 44751.973 &\phn$     -5.44 $&$     -0.19 $&   0.728 \nl
45915.875 &$        -22.36 $&\phs   0.19  &   0.415 && 47005.901 &$        -87.93 $&\phs   0.16  &   0.461 && 44781.895 &\phn$     -6.07 $&$     -0.69 $&   0.736 \nl
46040.610 &$        -25.32 $&$     -0.70 $&   0.476 && 47086.819 &$        -88.51 $&$     -0.05 $&   0.553 && 44870.773 &\phn$     -5.69 $&\phs   0.10  &   0.759 \nl
46334.784 &$        -27.54 $&\phs   1.49  &   0.618 && 47214.710 &$        -90.35 $&$     -0.03 $&   0.699 && 44935.742 &\phn$     -5.73 $&\phs   0.39  &   0.776 \nl
46657.855 &$        -31.03 $&$     -0.25 $&   0.775 && 47317.951 &$        -92.83 $&$     -0.20 $&   0.816 && 45069.031 &\phn$     -6.54 $&\phs   0.30  &   0.810 \nl
46704.694 &$        -30.56 $&\phs   0.01  &   0.798 && 47377.833 &$        -94.11 $&$     -0.21 $&   0.884 && 45583.852 &\phn$     -9.85 $&$     -0.29 $&   0.943 \nl
46717.666 &$        -30.81 $&$     -0.32 $&   0.804 && 47763.850 &$        -89.01 $&$     -0.22 $&   0.324 && 45667.793 &$        -10.46 $&$     -0.58 $&   0.965 \nl
46970.913 &$        -26.91 $&$     -0.10 $&   0.927 && 48162.793 &$        -92.27 $&$     -0.42 $&   0.778 && 45938.833 &$        -11.21 $&$     -0.76 $&   0.035 \nl
47005.822 &$        -25.67 $&\phs   0.38  &   0.943 && 48232.697 &$        -93.83 $&$     -0.38 $&   0.858 && 46377.843 &\phn$     -9.19 $&\phs   0.53  &   0.148 \nl
47317.945 &$        -20.06 $&$     -0.55 $&   0.095 && 49175.929 &$        -94.30 $&\phs   0.19  &   0.932 && 46657.892 &\phn$     -8.66 $&$     -0.18 $&   0.221 \nl
47763.838 &$        -19.50 $&$     -0.05 $&   0.311 && 49261.773 &$        -93.90 $&\phs   0.41  &   0.030 && 46704.702 &\phn$     -7.72 $&\phs   0.52  &   0.233 \nl
48162.783 &$        -25.62 $&$     -0.05 $&   0.504 && 49560.953 &$        -88.33 $&\phs   0.04  &   0.370 && 46717.714 &\phn$     -7.65 $&\phs   0.53  &   0.236 \nl
49175.900 &$        -23.78 $&$     -0.16 $&   0.995 && 50130.073 &$        -95.07 $&$     -0.64 $&   0.018 && 46970.947 &\phn$     -6.74 $&\phs   0.13  &   0.302 \nl
49560.915 &$        -18.08 $&$     -0.09 $&   0.181 && 50364.719 &$        -89.54 $&$     -0.26 $&   0.285 && 46980.968 &\phn$     -7.29 $&$     -0.47 $&   0.304 \nl
50364.707 &$        -28.14 $&$     -0.43 $&   0.571 && 50388.604 &$        -88.93 $&$     -0.01 $&   0.313 && 47005.907 &\phn$     -6.83 $&$     -0.14 $&   0.311 \nl
50388.599 &$        -28.43 $&$     -0.38 $&   0.582 &&           &                 &             &         && 47086.824 &\phn$     -6.81 $&$     -0.52 $&   0.331 \nl
          &                 &             &         &&  \multicolumn{4}{c}{\bf HD 216219   }               && 47317.956 &\phn$     -5.21 $&\phs   0.05  &   0.391 \nl
 \multicolumn{4}{c}{\bf HD 204613   }               && 42675.765 &\phd\phd$     -9.0^{\displaystyle*} $&\phs   0.02  &   0.192 && 47377.839 &\phn$     -5.03 $&\phs   0.00  &   0.407 \nl
43385.862 &$        -88.1^{\displaystyle*} $&\phs   0.53  &   0.339 && 42676.720 &\phd\phd$     -9.7^{\displaystyle*} $&$     -0.69 $&   0.192 && 47763.842 &\phn$     -4.25 $&$     -0.23 $&   0.506 \nl
45178.992 &$        -87.85 $&\phs   0.45  &   0.381 && 42677.855 &\phd\phd$     -9.2^{\displaystyle*} $&$     -0.19 $&   0.192 && 48162.788 &\phn$     -4.05 $&$     -0.09 $&   0.609 \nl
45314.625 &$        -87.91 $&\phs   0.43  &   0.535 && 44026.965 &\phn$     -4.02 $&$     -0.14 $&   0.541 && 48232.689 &\phn$     -4.04 $&\phs   0.02  &   0.628 \nl
45583.875 &$        -91.98 $&\phs   1.16  &   0.842 && 44073.898 &\phn$     -3.47 $&\phs   0.38  &   0.553 && 49175.923 &\phn$     -8.26 $&$     -0.09 $&   0.871 \nl
45667.801 &$        -94.73 $&$     -0.20 $&   0.937 && 44085.922 &\phn$     -3.95 $&$     -0.10 $&   0.556 && 49261.783 &\phn$     -7.99 $&\phs   0.65  &   0.893 \nl
45917.827 &$        -89.51 $&\phs   0.87  &   0.222 && 44110.844 &\phn$     -3.21 $&\phs   0.64  &   0.563 && 49560.963 &\phn$     -9.22 $&\phs   0.74  &   0.971 \nl
45938.838 &$        -90.61 $&$     -0.69 $&   0.246 && 44141.934 &\phn$     -3.64 $&\phs   0.21  &   0.571 && 50364.714 &\phn$     -8.68 $&\phs   0.57  &   0.178 \nl
46334.768 &$        -90.28 $&\phs   0.00  &   0.697 && 44173.801 &\phn$     -4.04 $&$     -0.18 $&   0.579 &&           &                 &             &         \nl
46657.877 &$        -93.26 $&\phs   0.52  &   0.065 && 44191.676 &\phn$     -3.43 $&\phs   0.44  &   0.584 &&           &                 &             &         \nl
\tablenotetext{{\displaystyle*}}{Velocities marked by an asterisk are from Luck and Bond (1991)}

\enddata
\end{deluxetable}

\clearpage

\begin{deluxetable}{l@{\hspace{5pt}}r@{\hspace{9pt}}rrlrl@{}rl}
\tablenum{2}
\tablewidth{0pt}
\tablecaption{Orbital Elements for sgCH Stars}
\tablehead{
\colhead{Star}             & \colhead{P}             &
\colhead{$\gamma$}         & \colhead{K}             &
\colhead{e}                & \colhead{$\omega$}      &
\colhead{T(JD)}            & \colhead{$a$ sin $i$}   &
\colhead{\phn f(m)} \\
\colhead{}                 & \colhead{(days)}        &
\colhead{(km s$^{-1}$)}    & \colhead{(km s$^{-1}$)} &
\colhead{}                 & \colhead{(deg)}        &
\colhead{(-2400000$^d$)} & \colhead{(Gm)}          &
\colhead{(M$_\odot$)}}

\startdata
HD 11377 & $4140\pm$118 & $-25.88\pm$0.27 &$3.88\pm$0.52 & $0.16\pm$0.08
& $65\pm$36 & $45240\pm$437 & $218\pm$30\phd & $0.0241\pm$0.0097 \nl
BD~$+17^\circ2537$&$1796\pm$12\phn&$1.13\pm$0.13 & $7.17\pm$0.22 & $0.14\pm$0.02
& $186\pm$24 & $46291\pm$128 & $175.3\pm$5.6 & $0.0668\pm$0.0063 \nl
HD 122202 & $1290\pm$9\phn\phn & $-11.15\pm$0.18 & $3.34\pm$0.23 & $0.09\pm$0.08
& $238\pm$62 & $46994\pm$205 & $59.0\pm$4.2 & $0.0049\pm$0.0010 \nl
HD 202020 & $2064\pm$10\phn & $-24.49\pm$0.14 & $6.45\pm$0.20 & $0.08\pm$0.04
& $279\pm$19 & $47122\pm$105 & $182.0\pm$5.6 & $0.0568\pm$0.0052 \nl
HD 204613 &$878\pm$4\phn\phn & $-90.96\pm$0.11 & $3.29\pm$0.15 & $0.13\pm$0.05
& $192\pm$21 & $47479\pm$49 & $39.4\pm$1.8 & $0.0032\pm$0.0004 \nl
HD 216219 & $3871\pm$39\phn & $-6.96\pm$0.08 & $3.31\pm$0.13 & $0.06\pm$0.03
& $159\pm$29 & $45803\pm$308 & $175.8\pm$7.1 & $0.0145\pm$0.0017 \nl

\enddata
\end{deluxetable}

\begin{figure}
\figurenum{1}
\plotfiddle{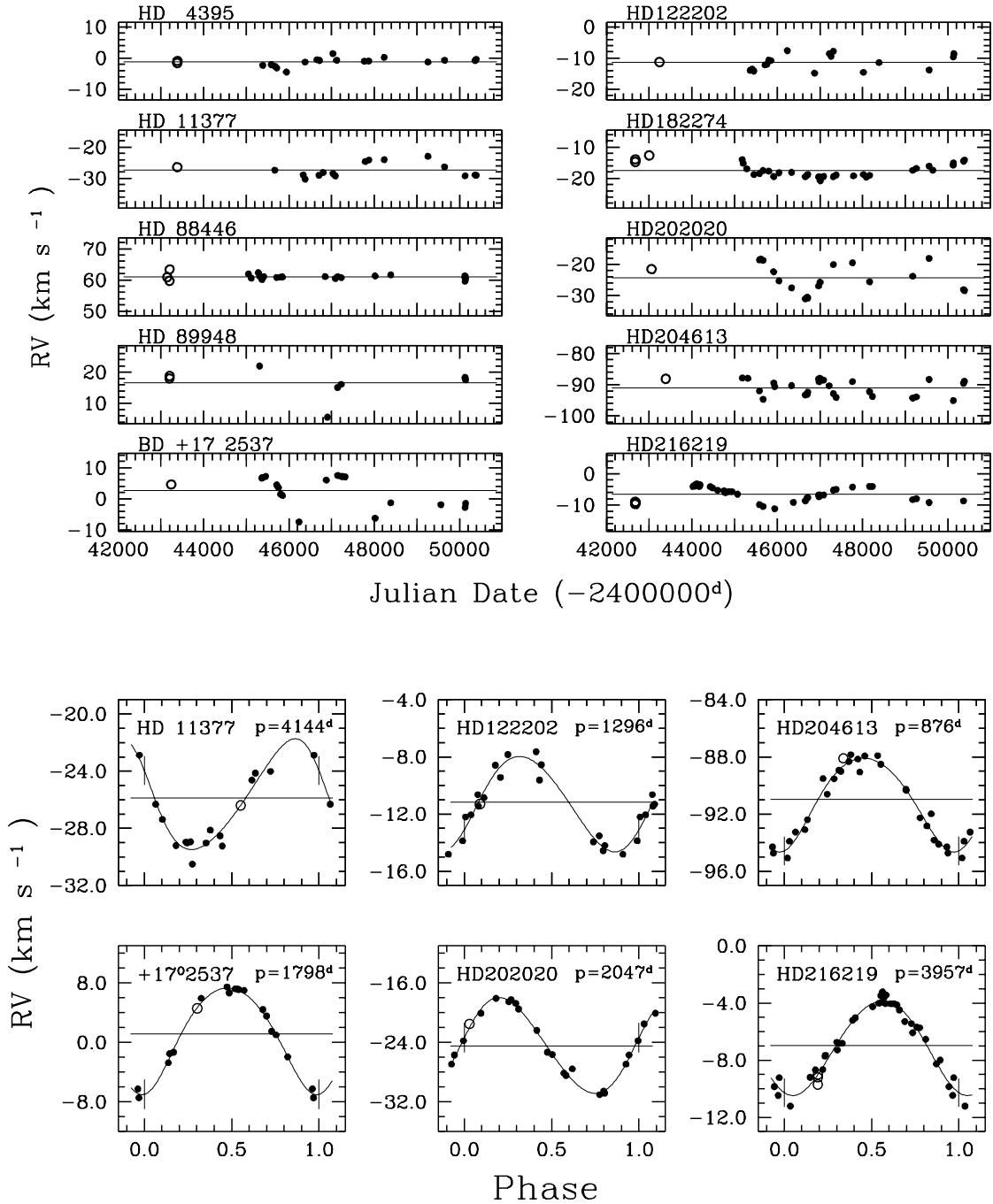}{7.1in}{0degrees}{90}{90}{-290pt}{-90pt}
\caption{Radial Velocities versus Julian dates for sgCH stars are
shown in the upper part of the figure.  In the lower part, Radial
Velocities versus Phase are shown for the computed orbits
(solid curves) from the elements listed in Table 2.  The observed
velocities from which these orbits were calculated are plotted as
dots for the present velocities, and open circles for the velocities
taken from Luck and Bond (1991).  \label{fig1}}
\end{figure}

\begin{figure}
\figurenum{2}
\plotfiddle{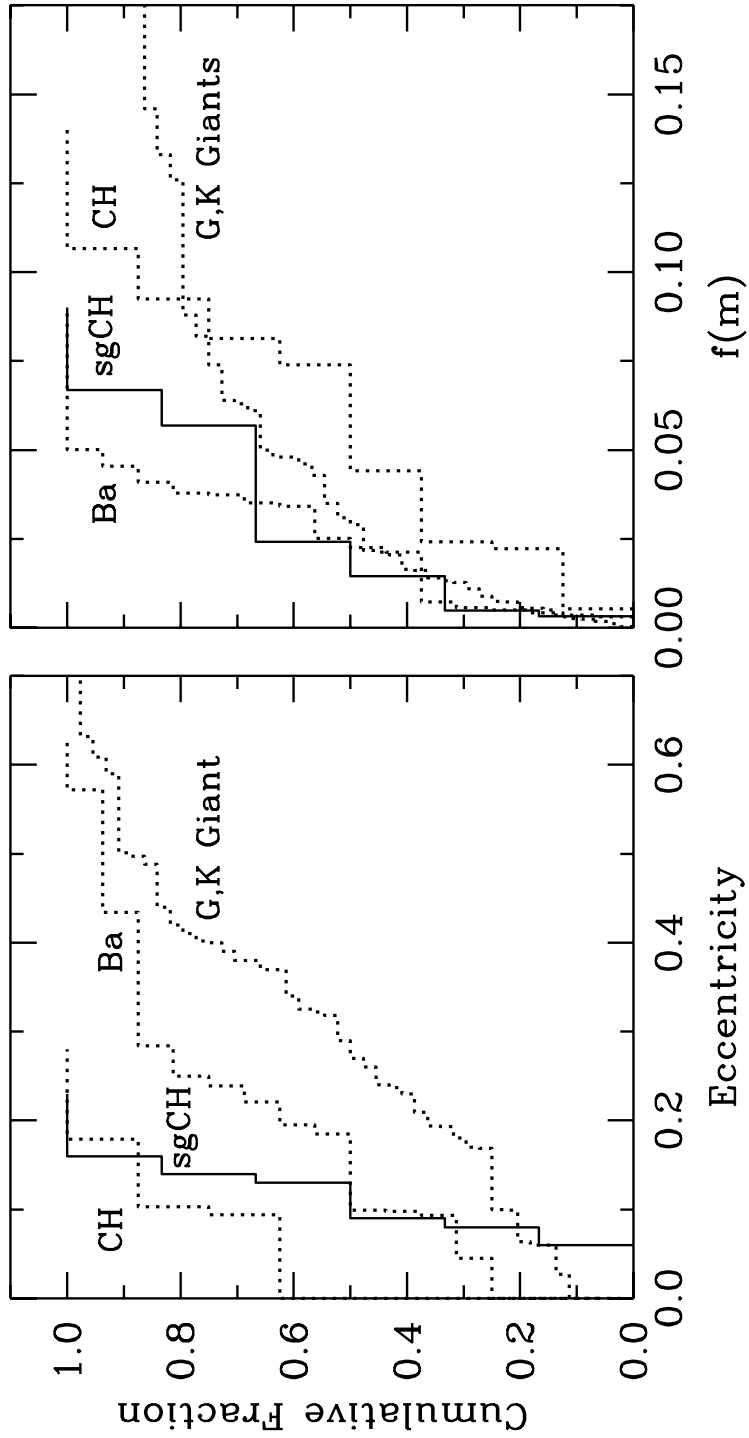}{7.1in}{0degrees}{90}{90}{-290pt}{-90pt}
\figcaption{On the left, the fraction of stars with eccentricities less
than a given eccentricity for sgCH stars (solid curve), CH stars,
barium stars and normal G, K giants (dotted curves).  On the right: the
same except for mass function rather than eccentricity.  \label{fig2}}
\end{figure}

\end{document}